\documentclass[aps,twocolumn,showpacs,superscriptaddress,amsmath]{revtex4}
\usepackage{amsmath,amscd,amssymb,euscript,eufrak}
\usepackage{graphicx}
\newcommand{\bea}{\begin{eqnarray}}
\newcommand{\beq}{\begin{equation}}
\newcommand{\eea}{\end{eqnarray}}
\newcommand{\eeq}{\end{equation}}

\begin{document}
\title{Triplet absorption spectroscopy and electromagnetically induced transparency}
\author{F. Ghafoor}
\affiliation{Department of Physics, COMSATS Institute of
Information Technology, Islamabad, Pakistan }
\author{R.G. Nazmitdinov}
\affiliation{Departament de F{\'\i}sica, Universitat de les Illes Balears, E-07122 Palma
de Mallorca, Spain}
\affiliation{Bogoliubov Laboratory of Theoretical Physics, Joint Institute for Nuclear
Research, 141980 Dubna, Russia}
\begin{abstract}
Coherence phenomena in four-level atomic system, cyclicly driven
by three coherent fields, are investigated  thoroughly at zero and
weak magnetic fields. Each strongly interacting atomic state is
converted to a triplet due to a dynamical Stark effect.
Two dark lines with a Fano-like profile are arising in the triplet
absorption spectrum with anomalous dispersions.
We provide the conditions to
control the widths of  the transparency windows by means
of the relative phase
of the driving fields and the intensity of the microwave field,
that closes the optical system loop.
The effect of the Doppler
broadening on results of the triplet absorption spectroscopy is
analysed in detail.
\end{abstract}
\date{\today}
\maketitle

\section{Introduction}

The absorption spectrum of two-level atomic system exhibits
Lorentzian line shape in the absence of any driving field
\cite{zub}. The spectrum is modified when the excited state is
coupled to another excited state by a strong laser field. As a
result, each excited state splits on two components. This is
so-called the Autler-Townes (AT) doublet \cite
{Autler,Fano,Agarwal001} that has been extensively studied in
context of spontaneous emission spectrum
\cite{Agassi,Zhu-Narducci,Pasp}, stimulated absorption
\cite{Ph,Bjor,Gray,Delsart,Fisk}, wave mixing \cite
{Yanp,Zhiq,Yig,Yanp01}, to name just a few. Indeed, the doublets
appear in various 3-level atomic or molecular systems interacting
with a strong laser field \cite{eit}. In absorption spectra of
such systems a dark line appears in the probe excitation signal
under the ideal conditions. That dark line, which is the essence
of Electromagnetically Induced Transparency (EIT)
\cite{eit,eit02}, is due to a quantum interference between two
alternative indistinguishable transition pathways created by the
coupling fields with internal states of a quantum system.
The EIT
is also the manifestation of the Fano-like interference.
This interference is characterised by the asymmetric line shape of a resonance spectrum,
which is created by various mechanisms in different quantum systems \cite{mir}.
In particular, this asymmetric line shape of resonances
has been predicted in atomic ionization due to
laser-induced continuum structure (for a review see \cite{prk}).
The possibility to altering losses of optical beams by strong laser fields
attracts researches to employ
the EIT based schemes, for example, to slow
the group velocity of a subluminal optical probe pulse transmitted by
optical media
(for a review see \cite{nov}) .

The phenomenon of the AT doublet absorption spectroscopy can be
further generalized when a multi-level atom is considered. It is
expected that the Fano-like interference will be present in this
case as well. Evidently, the multiplicative action of these
mechanisms in  multi-level atom under strong laser fields opens a
new avenue to explore various aspects of the absorption
cancellation absent in the AT doublet spectroscopy.
In fact,
a connection of a triplet spectroscopy with the EIT presents a
real challenge to the field of atomic spectroscopy \cite{Related}.
Note, that the ability to control, for example, the properties of
double EIT (which is absent in the case of 3-level systems):
two transparency windows convert to one and vise
versa,  - may be used as an optical switcher in nanophotonics.

To elucidate typical features
of a triplet absorption spectroscopy
we consider four-level atomic schemes with different radiative
decay mechanisms, cyclicly driven by three coherent fields. In
addition, our systems interact with the microwave field which
frequency is much smaller of those of the optical fields. It will
be shown that the variation of the intensities and phases of
optical and microwave fields enables to one to control the degree
of super- and sub-luminality in the transparency windows in the
absorption triplet spectroscopy.  The influence of the Doppler
broadening, important from experimental point of view, will be
taken into account in our consideration. In this case the mismatch
of the optical and  microwave frequencies will be analysed
thoroughly.

The structure of the paper is as follows. In Sec. II we introduce
our model. The Doppler broadening and the vector mismatch
are highlighted here as well. In Sec.III
we discuss various aspects of triplet absorption spectroscopy,
when there are multiple decay channels. Sec.IV is devoted to the
analysis of triplet absorption spectroscopy,
when there is only one decay channel. Main conclusions are
summarized in Sec.V.

\section{The Model}
\subsection{Stationary electrical susceptibility}
\begin{figure}[t]
\centering
\hspace*{1cm} \includegraphics[width=3.2in]{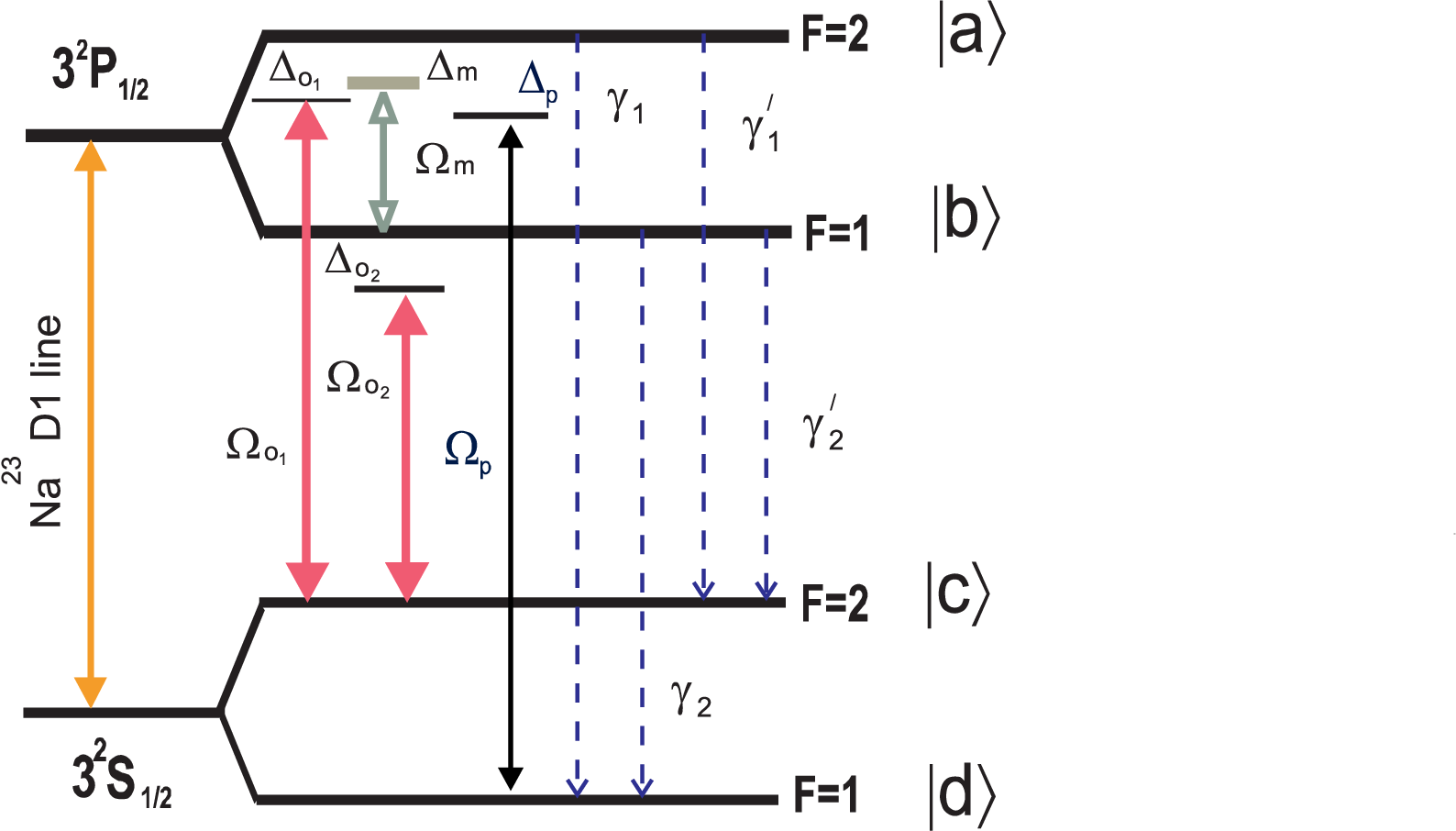}
\caption{(Color online) A sketch of the atomic energy spectrum of
Sodium D1 line at zero magnetic field. }
\label{f1}
\end{figure}

As a typical example of atomic system, interacting with optical
fields, we consider first the atomic scheme similar to the Sodium
$Na^{23}$ D1 line (3S$ _{1/2}^{2}\Longleftrightarrow$
3P$_{1/2}^{2}$) with $\lambda=5895.93{\textup{\AA}}$ at nonzero
spin-orbit interaction and zero magnetic field (see Fig.\ref{f1}).
Levels 3$^{2}$P$_{1/2},$F=2 ($\left\vert a\right\rangle$) and
3$^{2}$P$_{1/2},$F=1 ($\left\vert b \right\rangle $) of the
excited doublet are driven by a microwave field with the
Rabi-frequency $ \Omega _{m}$. These closely spaced levels are
coupled with the ground state 3 $^{2}$S$_{1/2},$F=2 ($\left\vert
c\right\rangle $) by means of two coherent optical fields with the
Rabi frequencies $\Omega_{o_1}$ and $\Omega _{o_2}$.

Excited states decay to the lowest ground state,
3$^{2}$S$ _{1/2},$F=2 ($\left\vert c\right\rangle $) and 3$^{2}$S$
_{1/2},$F=1 ($\left\vert d\right\rangle $) by means of allowed
electric dipole transitions. A weak probe field (with the Rabi
frequency $\Omega _{p}$) couples the lowest ground state
$\left\vert d\right\rangle $ and the excited state $\left\vert
a\right\rangle$. We define the optical field detuning parameters
as: $\Delta_{o_1}=\omega_{ca}-\omega_{o_1}$, $\Delta_{o_2}=
\omega_{cb}-\omega_{o_2}$. The detuning parameters of the
microwave and probe fields are $\Delta_m=\omega_{ba}-\omega_m$,
and $\Delta_p=\omega_{da}- \omega_p$, respectively. Here, the
frequencies $\omega_{o_1,o_2}$ are associated with two optical fields,
$\omega_m$ corresponds to the microwave field, while the frequency
$\omega_p$ characterises the probe field.

In general, the Rabi frequencies can be complex
$\Omega_i=|\Omega_i|\exp(i\phi_i)\,, i=o_1,o_2,m,p $.
For the sake of simplicity, we
consider the phase associated with the probe field to be fixed as
$\phi_p=0$. Below, for the sake of convenience, we omit the
modulus, i.e., $ |\Omega_i|\Rightarrow \Omega_i$.

The response of a medium due to interaction of these fields with the atomic
system under consideration can be found with the aid of the susceptibility
of the system (see for details \cite{zub})
\begin{equation}
\chi=\frac{N \wp_{ad}}{\epsilon_0 E} \rho_{ad}\,.
\end{equation}
Here, $\wp_{ad}$ is a dipole matrix element, $N$ is number density of atom
gas, and $\rho_{ad}$ is the density matrix element between states $|a\rangle$
and $|d\rangle$.

To trace a dynamical behaviour of the probe pulse in the medium we need to
evaluate the group index for our system. It is related to the group velocity
$v_g$ via $N_g$=c/$v_g$, where c is the speed of light in a vacuum. The
group index is calculated as:
\begin{equation}  \label{grin}
N_g=1+2\pi \mathrm{Re}[\chi]+2\pi\omega_{ad}\mathrm{Re}[\frac{\partial\chi}{%
\partial\Delta_p}].
\end{equation}
Thus, to analyse the response of our systems to external fields, we have to
know the time evolution of the density matrix $\rho_{ad}$. To solve that
problem we employ the master equation in Lindblad form
\begin{eqnarray}
\label{Leq}
\frac{d\rho}{dt}=-\frac{i}{\hbar}[H_I ,\rho]+\Lambda\rho\,,
\end{eqnarray}
with a damping term:
\begin{equation}
\label{dter}
\Lambda\rho=-\frac{1}{2}\Gamma \sum( \sigma^\dagger \sigma \rho+\rho
\sigma^\dagger \sigma-2\sigma \rho \sigma^\dagger).
\end{equation}
Here $\sigma=\left\vert a\right\rangle \left\langle d\right\vert $ or $%
\left\vert b\right\rangle \left\langle d\right\vert$ and $\sigma^{\dagger}
=\left\vert d\right\rangle \left\langle a\right\vert$ or $\left\vert
d\right\rangle \left\langle b\right\vert $ are the lowering and raising
operators for the two decay transitions of the system, respectively. In
particular, for the spontaneous decay rate of the excited state $%
3^{2}P_{3/2}\rightarrow 3^{2}S_{1/2}$ one has
\begin{equation}
\Gamma=\frac{1}{4\pi\varepsilon_0}\frac{4|\wp_{ad}|^2\omega_{da}^3}{3\hbar
c^3}\equiv\gamma\,.
\end{equation}
Our system is described by the Hamiltonian, taken in the
interaction representation and in the rotating wave approximation
(see for details of derivation, for example, \cite{zub}):

\begin{eqnarray}
\label{int}
 H_{I} &
=-\frac{\hbar}{2}[\Omega_{m}e^{i(\Delta_{m}t+\varphi_m)}\left\vert
a\right\rangle \left\langle b\right\vert +\Omega_{o_{1}}e^{i(
\Delta_{o_{1}}t+\varphi_{o_1})}\left\vert a\right\rangle
\left\langle c\right\vert \nonumber
\\
& +\Omega_{o_{2}}e^{i(\Delta_{o_{2}}t+\varphi_{o_2})}\left\vert
b\right\rangle \left\langle c\right\vert +\Omega_{p}e^{i
\Delta_{p}t}\left\vert a\right\rangle \left\langle d\right\vert]
+H.c.
\end{eqnarray}

To proceed further, we use the transformations
\begin{eqnarray}
\label{trs}
&&\rho_{ad} \rightarrow
\tilde{\rho}_{ad}e^{i\Delta_{p}t}\,,\nonumber\\
&&\rho_{cd} \rightarrow
\tilde{\rho}_{cd}e^{-i(\Delta_{o_1}-\Delta _{p})t}\,,\nonumber\\
&&\rho_{bd} \rightarrow
\tilde{\rho}_{bd}e^{-i(\Delta_{m}-\Delta _{p})t}\,,\\
&&\rho_{ba}\rightarrow
\tilde{\rho}_{ba}e^{-i\Delta_{m}t}\,,\nonumber\\
&&\rho_{cd} \rightarrow\tilde{\rho}_{cd}e^{-i\Delta_{o_1}t}\,.\nonumber
\end{eqnarray}

By means of (\ref{Leq},\ref{dter}), in virtue of the transformation (\ref{trs}),
one obtains the following three coupled rate equations
for slowly varying amplitudes (in the first order of the probe
field and all orders of the coherent driving fields)

\begin{eqnarray}
\frac{d\tilde{\rho}_{ad}}{dt}& = &-[i\Delta_{p}+\frac{1}{2}
(\gamma_{1}+\gamma^\prime_{1})]\tilde{\rho}_{ad}+\frac{i}{2}
\Omega_{m}e^{i\varphi_m}\tilde{\rho}_{bd} + \nonumber\\
&+&\frac{i}{2}\Omega_{o_{1}}e^{i\varphi_{o_1}}\tilde{\rho}_{cd}-\frac{i}{2}\Omega
_{p}( \tilde{\rho}_{aa}-\tilde{\rho}_{dd})\,,\\
\frac{d\tilde{\rho}_{bd}}{dt}& = &-[i(\Delta_{p}-\Delta_{m})+
\frac{1}{2}(\gamma_{2}+\gamma^\prime_{2})]\tilde{\rho}_{bd}+
 \nonumber\\
& + &\frac{i}{2}\Omega^\star_{m}e^{-i\varphi_m}\tilde{\rho}_{ad} +
\frac{i}{2}\Omega_{o_{2}}e^{i\varphi_{o_2}}\tilde{\rho}_{cd}-\frac{i}{2}\Omega _{p}\tilde{\rho}_{ba}\,,\\
\frac{d\tilde{\rho}_{cd}}{dt}& = &-i(\Delta_{p}-\Delta _{o_1})
\tilde{\rho}_{cd}+\frac{i}{2}\Omega_{o_{1}}e^{i\varphi_{o_1}}\tilde{\rho }_{ad} +\nonumber\\
& +
&\frac{i}{2}\Omega_{o_{2}}e^{i\varphi_{o_2}}\tilde{\rho}_{bd}-\frac{i}{2}\Omega
_{p} \tilde{\rho}_{ca}.
\end{eqnarray}
It is natural to employ the following initial state conditions:
$\tilde{\rho }_{dd}=1$, $\tilde{\rho}_{aa}=0$,
$\tilde{\rho }_{ba}=0$, $\tilde{ \rho}_{ca}=0$.
The above set of equations can be solved for $\tilde{\rho }_{ad}$ with the aid of the equation
\begin{equation}
R(t)=\int_{-\infty }^{t}e^{-Q(t-t^{\prime})}Pdt^{\prime}=Q^{-1}P,
\end{equation}
where $R(t)$ and $P$ are column matrices, while Q is a 3x3 matrix.

The solution is:
\begin{equation}
\tilde{\rho }_{ad}=-\mathrm{Re}[\tilde{\rho}_{ad}]-i\mathrm{Im}[\tilde{\rho }_{ad}]
\end{equation}
with
\begin{eqnarray}
\label{rs}
\mathrm{Re}[\tilde{\rho }_{ad}] &=&\frac{2(\gamma
_{2}+\gamma _{2}^{\prime })B_G(\Delta _{o_{1}}-\Delta
_{p})+R_GA_G}{A_G^{2}+B_G^{2}}
\Omega _{p}\,, \\
\label{is}
\mathrm{Im}[\tilde{\rho }_{ad}] &=&\frac{2(\gamma
_{2}+\gamma _{2}^{\prime })A_G(\Delta _{o_{1}}-\Delta
_{p})-R_GB_G}{A_G^{2}+B_G^{2}}\Omega _{p}\,.
\end{eqnarray}
Here, we use the following notations:
\begin{eqnarray}
\label{AG0}
A_{G} &  =-2(\Delta_{m}-\Delta_{p})[(\gamma_{1}+\gamma_{1}^{\prime}%
)(\gamma_{2}+\gamma_{2}^{\prime})\nonumber\\
&  +4\Delta_{o_{1}}\Delta_{p}-4\Delta_{p}^{2}+\Omega_{m}^{2}]-2\Delta_{o_{1}%
}\Omega_{o_{1}}^{2}\\
&  +2\Delta_{p}(\Omega_{o_{1}}^{2}+\Omega_{o_{2}}^{2})+2\Omega_{o_{1}}%
\Omega_{o_{2}}\Omega_{m}\cos\varphi,\,\nonumber\\
\label{BG0}
B_{G} &  =(\gamma_{1}+\gamma_{1}^{\prime})[4(\Delta_{p}-\Delta_{m}%
)(\Delta_{o_{1}}-\Delta_{p})+\Omega_{o_{2}}^{2}]\nonumber\\
&  +(\gamma_{2}+\gamma_{2}^{\prime})[4(\Delta_{m}-\Delta_{p})\Delta_{p}%
+\Omega_{o_{1}}^{2}]\,,\\
\label{RG0} R_{G} &
=4(\Delta_{o_{1}}-\Delta_{p})(\Delta_{p}-\Delta_{m})+\Omega_{o_2}^{2}\,,
\end{eqnarray}
and introduce a relative phase $\varphi=\varphi_{o_1}+\varphi_{o_2}-\varphi_m$.

Thus, we have derived  general expressions for the real and imaginary parts of the density matrix
$\tilde{\rho }_{ad}$, that allow us to analyse optical properties of
an ideal four-level atomic scheme displayed on Fig.\ref{f1}.

\subsection{The Doppler broadening}

To model an experimental situation we have to
take into account the random motion of atoms
due to thermal energy. Thermal atomic motion produces
a spreading of the absorbed frequency. It results
in the broadening of the optical profiles, so-called the
Doppler broadening. As a result, for each
field "i" the detuning $\Delta_i$ should be replaced by
$\Delta_i + k_i v$, where $k_i= \omega_i/c$ the wavevector of that field.
We assume, however, that optical fields have similar transition
frequencies
\begin{equation}
k\approx k_{o_1}\approx k_{o_2} \approx k_{p}\,.
\end{equation}
Below we consider a case, when
the optical, the microwave and the probe fields propagate collinearly
along the $+z$ direction.  The microwave frequency is much smaller
then those of other driving fields.
Therefore, the wave vector of the
microwave field  can be synchronized with the
wave vectors of the optical fields through the two-photon
resonance condition as
\begin{equation}
\Delta _{m}\approx \lbrack (\Delta _{o_{1}}-\Delta _{o_{2}})+\beta
kv],
\end{equation}
where $\beta\approx k_m/k$.

Taking into account the above conditions that lead to the
velocity dependent terms: $\Delta _{m}\rightarrow (\Delta _{m}+
\beta kv)$, $\Delta _{o_{1}}\rightarrow (\Delta _{o_{1}}+ kv)$, $
\Delta _{o_{2}}\rightarrow (\Delta _{o_{2}}+kv)$ and $\Delta
_{p}\rightarrow (\Delta _{p}+kv)$, we obtain the following form for the
susceptibility
\begin{equation}
\chi (kv)=\frac{N|\wp _{ad}|^{2}}{\epsilon _{0}\hbar \Omega _{p}}
[M(kv)\Omega _{p}]\,,
\end{equation}%
where
\begin{eqnarray}
&&M(kv)=-\frac{M_1(kv)+M_2(kv)+M_3(kv)}{A_{G_{D}}^{2}(kv)+B_{G_{D}}^{2}(kv)}\,,\\
&&M_1(kv)=R_{G_{D}}(kv)A_{G_{D}}(kv)\,,\nonumber\\
&&M_2(kv)=2(\gamma _{2}+\gamma_{2}^{\prime})G_{G_{D}}(kv)B_{G_{D}}(kv)\,,\nonumber\\
&&M_2(kv)=iQ(kv)\nonumber\,.
\end{eqnarray}
Here,
\begin{eqnarray}
&&Q(kv) =2A_{G_{D}}(kv)G(kv)[(\gamma _{2}+\gamma _{2}^{\prime
})-(\Delta_{p}+kv)]  \nonumber \\
&&-R_{G_{D}}(kv)B_{G_{D}}(kv)\,,\\
&&G_{G_{D}}(kv) = [\Delta _{o_{1}}-\Delta _{p}]\,,
\end{eqnarray}
and we have obtained the following expressions
\begin{eqnarray}
A_{G_{D}} &  =-2[(\Delta_{m}+\beta
kv)-(\Delta_{p}+kv)]\times\lbrack
(\gamma_{1}+\gamma_{1}^{\prime})(\gamma_{2}+\gamma_{2}^{\prime})\nonumber\\
&  +4(\Delta_{o_{1}}+kv)(\Delta_{p}+kv)-4(\Delta_{p}+kv)^{2}+\Omega_{m}%
^{2}]\nonumber\\
&  -2(\Delta_{o_{1}}+kv)\Omega_{o_{1}}^{2}+2(\Delta_{p}+kv)(\Omega_{o_{1}}%
^{2}+\Omega_{o_{2}}^{2})\nonumber\\
&  +2\Omega_{o_{1}}\Omega_{o_{2}}\Omega_{m}\cos\varphi\,,\\
B_{G_{D}} &  =4[(\Delta_{p}+kv)-(\Delta_{m}+\beta kv)]\nonumber\\
&
\times\lbrack(\gamma_{1}+\gamma_{1}^{\prime})(\Delta_{o_{1}}-\Delta
_{p})+(\gamma_{2}+\gamma_{2}^{\prime})(\Delta_{p}+kv)]\nonumber\\
&
+(\gamma_{1}+\gamma_{1}^{\prime})\Omega_{o_{2}}^{2}+(\gamma_{2}+\gamma
_{2}^{\prime})\Omega_{o_{1}}^{2}\,,\\
R_{G_{D}} &
=4(\Delta_{o_{1}}-\Delta_{p})[(\Delta_{p}+kv)-(\Delta_{m}+\beta
kv)]+\Omega_{2}^{2}.
\end{eqnarray}

Thus, we obtain for the average susceptibility
\begin{equation}
\chi ^{(d)}=\frac{N|\wp _{ad}|^{2}}{\epsilon _{0}\hbar
V_{D}\sqrt{\pi }} \int_{-\infty }^{\infty
}M(kv)e^{-\frac{(kv)^{2}}{V_{D}^{2}}}d(kv)\,,
\end{equation}
which is defined with the aid of the Maxwell-Boltzman distribution,
where
$V_{D}=\sqrt{k_{B}T\omega ^{2}/Mc^{2}} $ is the Doppler width. The
Doppler width is a free parameter in our numerical examples. The
corresponding group index $N_{g}^{(d)}$ has the form (\ref{grin}),
where $\chi $ is replaced by $\chi ^{(d)}$. Although we present a
general scheme, below a thorough analysis is provided for a resonant
interaction as an example, i.e., $\Delta _{i}=0,i=o_{1},o_{2},m$.

\section{Triplet absorption spectroscopy}
To proceed further we separate the susceptibility $\chi$ on
the imaginary $\Im m \chi\equiv\chi_{\Im}$ and real
$\Re e\chi\equiv\chi_{\Re}$ parts
\begin{eqnarray}
\label{ic}
\chi_{\Re}&=&-\frac{N|\wp_{ad}|^2}{\epsilon_0\hbar\Omega_p}\mathrm{Re}[
\tilde{\rho}_{ad}]\,, \\
\chi_{\Im}&=&-\frac{N|\wp_{ad}|^2}{\epsilon_0\hbar\Omega_p}\mathrm{Im}[
\tilde{\rho}_{ad}]\,.
\end{eqnarray}
It is convenient to measure the absorption and dispersion in the
units of $ \frac{N|\wp_{ad}|^2}{\epsilon_0\hbar\Omega_p}$. Most
BEC experiments reach quantum degeneracy between 500 nK and 2
$\mu$K, at densities between $ 10^{14} $ and $10^{15}$ cm$^{-3}$
\cite{ket}. For the number density $ N\approx1.3\times10^{14}$
cm$^{-3}$, $\wp_{ad}=2.492\times ea_0$ in case of the Sodium D1
line ($a_0$ is the Bohr radius) the pre-factor $\frac{
N|\wp_{ad}|^2}{\epsilon_0\hbar}$ is about $2\pi\times 9.89$ MHz.
This value could be associated with the spontaneous decay rate
$\Gamma$ of the Sodium D1 line.

In the analysis below we use $\Gamma\equiv\gamma_1=\gamma$ as a
natural unit of the relevant physical quantities and consider Rabi
frequency of the weak probe field as $\Omega_p=0.1\gamma$ when
$\wp_{ad}=2.492\times ea_0$ is chosen for the atomic Sodium D1
line. Below some representative examples for the Doppler-free and
the Doppler-broadened system are discussed in detail.
In this section we consider the value
$\varphi =\pi/2$ for the relative phase of the driving fields.

\begin{figure*}
\centering
\includegraphics[width=5.5 in]{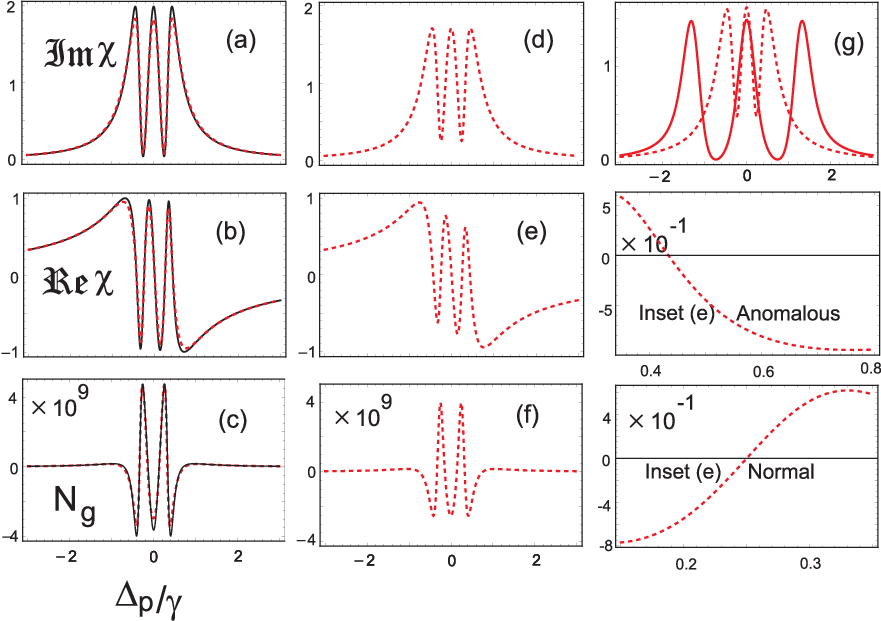}
\caption{(Color online) Absorption, dispersion, and group index,
vs dimensionless probe field detuning $\Delta_p/\protect\gamma$.
The results  at equal intensities
$\Omega_{o_1}=\Omega_{o_2}=\Omega_{m}=0.5\protect\gamma$
(without  and with the Doppler broadening, see panels (a), (b), (c))
are connected by solid (black) and dotted (red) lines,
respectively. The results  at the same intensities (with the
Doppler broadening and the vector mismatch $\beta=0.5$, see panels (d),(e), (f))
are connected by dotted (red) line.
The results at the vector mismatch $\beta=0.5$ and
at equal intensities values: $0.5\gamma$ and  $1.5\gamma$ are
connected by dotted (red) and solid (red) lines, respectively
(see panel (g)).
On two right bottom panels the insets  for the case (e) are displayed.
The following parameters are used:
$\protect\omega=10^8\protect\gamma$,
$V_D=0.2\protect\gamma$, $\protect \varphi=\protect\pi/2$,
$\gamma_1^\prime=\gamma_2=\gamma_2\prime=0.01\gamma$,
}
\label{f5}
\end{figure*}
\begin{figure*}[tbp]
\centering
\includegraphics[width=5in]{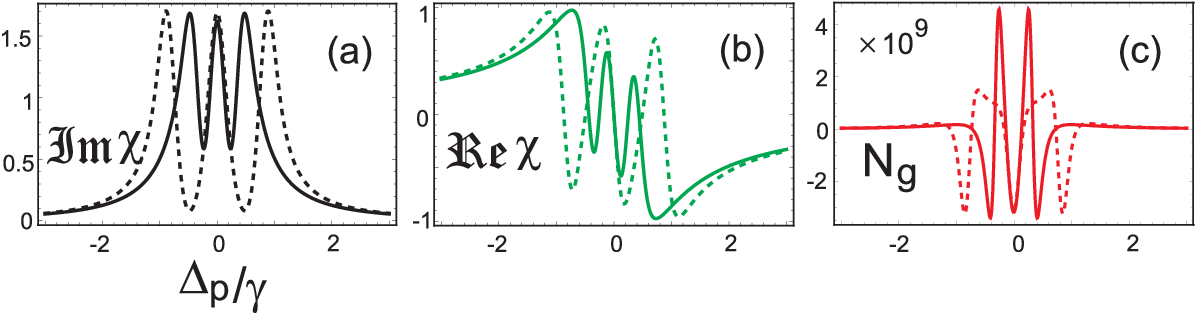}
\caption{(Color online)
Absorption (a), dispersion (b), and
group index (c) vs dimensionless probe field detuning
$\Delta_p/\protect\gamma$ at $\beta=0.5$.
The results for
$\Omega_{o_1}=\Omega_{o_2}=\Omega_{m}=0.5\protect\gamma$, and $=\protect\gamma$
are connected by solid and dash lines, respectively.
The other parameters are kept the same as in Fig. 2.}
\label{f2a}
\end{figure*}

\subsection{
$\Omega_{o_1}=\Omega_{o_2}=\Omega_{m}\approx
\gamma_1\gg\gamma_2, \gamma_{1,2}^\prime$
}

To begin with, we analyze the case, that has been studied by many
authors in systems with
an excited doublet. For example, Knight and co-workers  \cite{Pasp}
analyzed a well known scheme of a quantum beat laser in context of
a spontaneous emission, using the same limit.

Generally, when the intensities of all driving
fields are kept comparable to the decay rates $\gamma_{1,2}$,
the absorption spectrum splits into a triplet
\cite{tapali,scully-knight}. Note that the triplet is associated
with a phenomenon of dynamical Stark triplet splitting.
Traditionally it is known as the Autler-Townes splitting.

The positions of the dark lines are determined from the
requirement that the numerator of
(\ref{is}) is equal zero. As a result, one obtains the following estimation
\begin{equation}
\Delta _{p}\approx \pm\frac{1}{2} \sqrt{\Omega
_{o_{2}}^{2}+\gamma_2^{2}}\,.
\end{equation}
These dark lines are
due to the quantum interference among three alternative
indistinguishable transition pathways to the excited state
$|a\rangle$. Qualitatively, this result can be understood from the
analysis of dressed states of three bare pathways: $|d\rangle
\rightarrow |a\rangle$, $|d\rangle \rightarrow
|a\rangle\rightarrow |b\rangle \rightarrow |a\rangle$, $|d\rangle
\rightarrow |a\rangle\rightarrow |c\rangle \rightarrow |a\rangle$,
coupled by two strong optical fields, and the microwave field.
However, this analysis requires a separate study and is beyond the
scope of the present paper.
\begin{figure*}[tbp]
\centering
\includegraphics[width=4in]{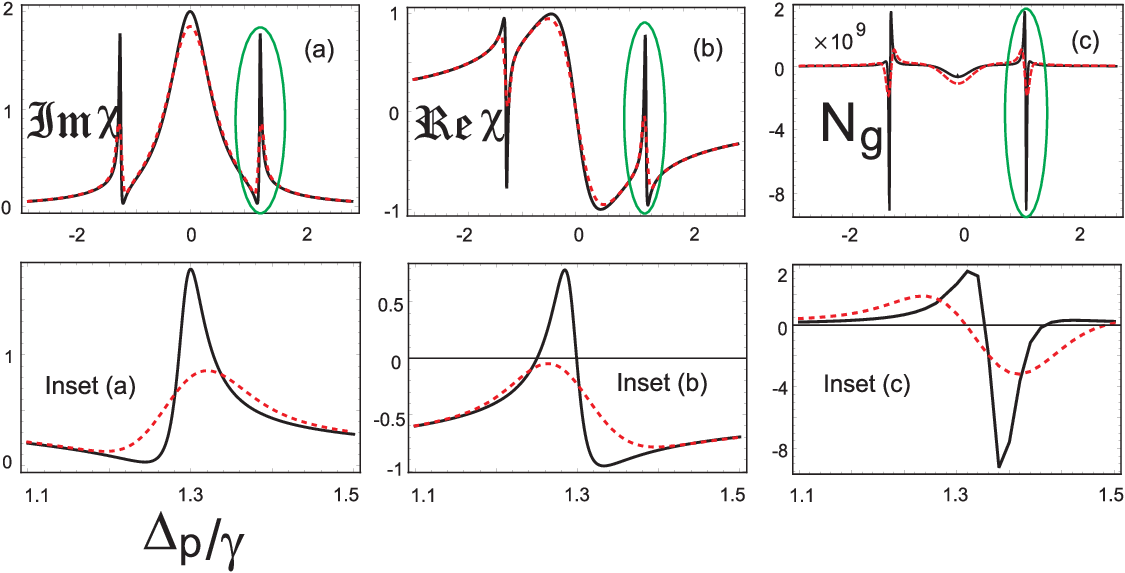}
\caption{(Color online) Top row: absorption (a), dispersion (b), and
group index (c) vs dimensionless probe field detuning
$\Delta_p/\protect\gamma$.
Bottom row: the insets for the case (a)-(c) are displayed.
The following parameters are chosen:
$\Omega_{o_1}= \Omega_{m}=0.5\protect\gamma$ and $
\Omega_{o_2}=2.5\protect\gamma$.
The results at $\beta=0,0.5$ are connected with
solid (black) and dashed (red) lines, respectively. The other
parameters are kept the same as in Fig.2.}
\label{f3}
\end{figure*}

 The absorption resonance positions
are defined from the requirement $A_G=0$ (see (\ref{AG0}))
 that yields three roots. Two of them
\begin{equation}
 \Delta _{p}\approx\pm\frac{1}{2} \sqrt{(\Omega
_{o_1}^{2}+\Omega _{o_2}^{2}+\Omega _{m}^{2})+\gamma _{1}
(\gamma_{2}+\gamma _{2}^{\prime })}
\label{sr}
\end{equation}
determine the positions of two side peaks, while
the central peak is located at $\Delta _{p}=0$.

Subsequently, it is easy to estimate each
peak height, using the corresponding root.
The  peak width is evaluated as a half
of the absorption maximum
\begin{equation}
\frac{\chi_{\Im}}{2}=-\frac{N|\wp_{ad}|^{2}}{B_G^{2}\epsilon_{0}\hbar}
[(\gamma _{2}+\gamma
_{2}^{\prime })A_G(\Delta _{o_{1}}-\Delta
_{p})-R_GB_G)]\,.
\label{wr}
\end{equation}
at the corresponding root
for $\Delta_p$ ($\Delta _{p}=0$, and see (\ref{sr})).

The quantum interference
 generates two dark lines that split the absorption spectra
into triplets (see Figs.2a,d,g).
 At equal intensities of the
optical and microwave fields, the widths of each
components  approach the value $\gamma /3$.
 As compared with the
traditional $\Lambda$ and $\Xi$ types of systems, here, in
addition to the sub-luminal light, the super-luminal light pulse
propagates through the three anomalous dispersive windows with
some absorption.

Thus, at equal intensities of the driving fields, there is
a dominance of the sub-luminal light
over the super-luminal one (see Fig.2c).
This dominance can be explained in
simple terms. In the insets of Fig.2e we zoom
the right wing of the dispersion ${\Re e}\chi$. Evidently, it is related to
the absorption spectrum via the Kramers-Kronig relation.
While the anomalous dispersion  decreases exponentially
with the increase of the ratio $\Delta_p/\gamma$,
the normal dispersion grows  faster with the increase of this ratio.

The sub-luminality and super-luminality are affected
by the vector mismatch. The vector mismatch
$\beta k$ degrades the interference; especially,
the super-luminality (compare Figs.2c,f).
The increase of the intensity increases the
EIT windows (see Fig.2g). However, this increase
yields: i)the decrease of the steepness of the
transparency windows; and  ii) the
widening of the absorption spectrum.
In addition, it leads to the attenuation of the
sub-liminality of the system (see Fig.\ref{f2a}).
Note, however, that the super-luminality is much less affected.

\subsection{$\Omega_{o_2}> \Omega_{o_1}=\Omega_{m},
\gamma_1\gg \gamma_2,\gamma_{1,2}^\prime$}

At the Doppler free case, it is convenient to estimate analytically
the locations of: i) two dark lines; ii) three
spectral components; and iii) their widths.
By means of the same recipe (see Sec.IIIA), we obtain for
the dark line positions
 \begin{equation}
\Delta_p\approx\pm\frac{1}{2}\sqrt{ \Omega_{o_2}^2+\gamma_2^2}\,.
\end{equation}
In this case the locations of side absorption resonances are determined by
(\ref{sr}). Similar to the results of Sec.IIIA, the central peak is located at
$\Delta _{p}=0$. In the considered conditions, (\ref{wr}) can be still useful
to define the absorption resonance widths.
Note that these
estimations agree with the corresponding results obtained
numerically in the presence of the Doppler broadening and the vector
mismatch.

The dominance one of the optical fields with respect to other
driving fields leads to the broadening of the central peak, while
the side resonances widths are decreasing. In this case the quantum
interference of three resonances produce a pattern that is a
typical for the Fano-like interference phenomenon \cite{mir}. The
Lorentzian shape of the central peak is complemented by two side
resonances with the asymmetric shapes (see Fig.\ref{f3}a).

The super-luminal behavior of the pulse around the two side-windows
with the anomalous dispersions is significantly enhanced, when the
intensity of the optical field $\Omega_{o_2}$ relative to other
driving fields is kept high (see Fig.\ref{f3}c).
On the other hand, the sub-luminality
at two side windows with a normal dispersion get worsened.
The largeness and smallness in the positive (negative) group index
is correlated through steepness and flatness in the normal
(anomalous) dispersion (compare Figs.\ref{f5}(e) and \ref{f3}(b)).

Thus, the vector mismatch of the microwave field with the other driving fields
degrades the interference. The larger is the value of the parameter
$\beta $, the larger is the degradation.
Although the vector mismatch attenuates strongly the
 super-luminality in the side transparency windows,
 the attenuation is much weaker in the central
transparency window (see Fig.\ref{f3}(a)).
At large strength of the optical field
$\Omega_{o_2}$ the rate of the atomic oscillations
between the second ground state and the excited state is fast as
compared with the decay rates of the states of excited doublet.
Therefore, the quantum interference is more effective due to the
decrease in the decay rates. However, the
inherited incoherence, produced by the vector mismatch of the microwave
field, may require the larger optical field intensity for its
suppression.
\begin{figure}[t]
\centering
\includegraphics[width=2.5in]{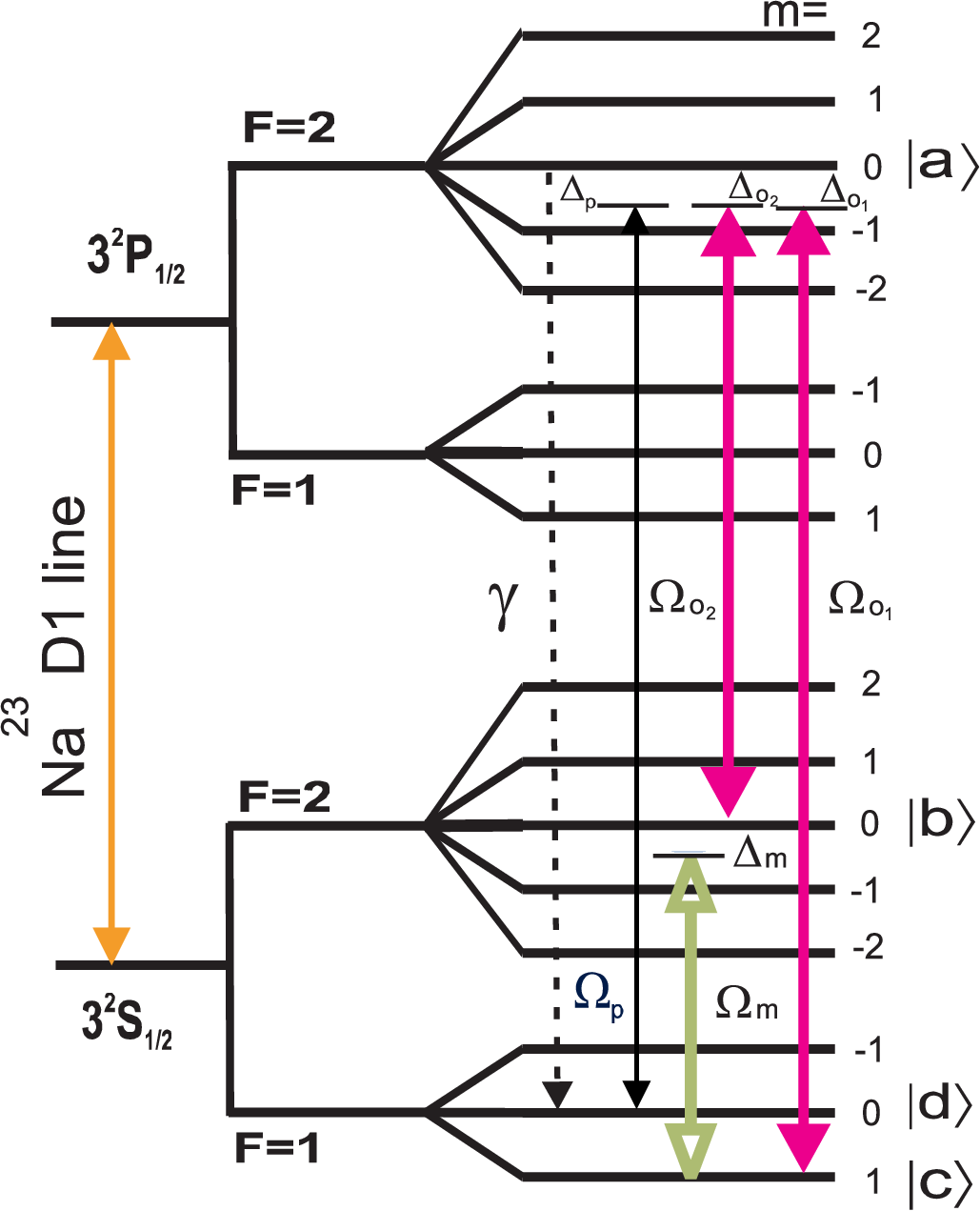}
\caption{ (Color online) A sketch of the atomic energy spectrum of Sodium D1
line at a weak magnetic field. }
\label{f2}
\end{figure}

\section{Simplistic triplet absorption spectroscopy}
\subsection{One decay channel}

The question arises: is it possible to simulate a physical situation, where
incoherent processes, discussed above, can be avoided ?
Fortunately, this can happen
in the scheme based on a simple probability loss,
when the contribution of decoherence due to the vector
mismatch is relaxed.

Let us consider the Sodium atom in a weak static magnetic field,
when there is only one decay channel. Evidently, the Zeeman splitting
yields the families of states
with different magnetic quantum numbers $m$
(see Fig.\ref{f2}). The microwave field with frequency
$\omega_m$ couples a low-lying state $|b\rangle$ and a ground
state $ |c\rangle$ with the Rabi frequency $\Omega_m$. These two
states are coupled with the excited state $|a\rangle$ by the two
optical fields of frequencies $ \omega_{o_1}$ and $\omega_{o_2}$
with the corresponding Rabi frequencies $ \Omega_{o_1}$ and
$\Omega_{o_2}$, to form a loop. A weak probe field couples the
excited state $|a\rangle$ with a low-lying state $|d\rangle$. The
simplicity in the radiative decay process in the present scheme
(observed experimentally \cite{exp})
provides the proper ground to study the interplay between the
driving fields at small decoherence background created by
the one decay channel.

This simplistic system is described by the Hamiltonian (taken in
the interaction representation and in the rotating wave
approximation)

\begin{eqnarray}
\label{sc1} H_{I} & =
&-\frac{\hbar}{2}[\Omega_{m}e^{i(\Delta_{m}t+\varphi_m)}\left\vert
b\right\rangle \left\langle c\right\vert +\Omega_{o_{1}}e^{i
(\Delta_{o_{1}}t+\varphi_{o_1})}\left\vert
a\right\rangle \left\langle c\right\vert  \nonumber\\
& +&\Omega_{0_2}e^{i(\Delta_{o_{2}}t+\varphi_{o_2})}\left\vert
a\right\rangle \left\langle b\right\vert
+\Omega_{p}e^{i\Delta_{p}t}\left \vert a\right\rangle \left\langle
d\right\vert] +H.c.
\end{eqnarray}

Applying our approach (see Sec.II) to the Hamiltonian (\ref{sc1}),
we obtain for the real and imaginary parts of the susceptibility
 \begin{equation}
\label{srs}
\chi_{\Re}=-\frac{N|\wp_{ad}|^{2}}{\epsilon_{0}\hbar\Omega_{p}}[\frac
{R_SA_S }{A_S^{2}+B_S^{2}}\Omega_{p}]\,,
\end{equation}
 and
\begin{equation}
\label{imsc}
\chi_{\Im}=+\frac{N|\wp_{ad}|^{2}}{\epsilon_{0}\hbar\Omega_{p}}[\frac
{ R_SB_S }{A_S^{2}+B_S^{2}}\Omega_{p}]\,.
\end{equation}
 Here
\begin{eqnarray}
\label{scabr}
A_{S} &  =2(\Delta_{o_{2}}-\Delta_{p})[4\Delta_{p}(\Delta_{p}-\Delta_{o_{1}%
})-\Omega_{o_{2}}^{2}]-2\Delta_{o_{1}}\Omega_{o_{1}}^{2}\nonumber\\
&  +2\Delta_{p}(\Omega_{o_{1}}^{2}+\Omega_{m}^{2})+2\Omega_{o_{1}}%
\Omega_{o_{2}}\Omega_{m}\cos\varphi\,,\\
B_{S} &  =4\gamma(\Delta_{p}-\Delta_{o_{2}})(\Delta_{o_{1}}-\Delta_{p}%
)+\gamma\Omega_{m}^{2}\,,\nonumber\\
R_{S} &  =4(\Delta_{o_{1}}-\Delta_{p})(\Delta_{p}-\Delta_{o_{2}}%
)+\Omega_{m}^{2}\,,\nonumber
\end{eqnarray}
where $\varphi=\varphi_m+\varphi_{o_1}-\varphi_{o_2}$

Note, that for the Doppler-free system, at the
resonance condition $\Delta_{o_1}=\Delta_{o_2}=0$, the absorption
is cancelled  if the equation
\begin{equation}
\label{pdl}
\Delta_p=\pm\frac{1}{2}\Omega_{m}
\end{equation}
is fulfilled, since $B_S=0$. Two dark lines are developed in the
Lorentzian type absorption spectrum irrespective of the decay rate
and the intensities of the optical fields. However, it is controlled
by the intensity of the microwave field. The positions of the
central absorption and the sides absorption peaks (components) of
the triplet are located at the roots $\Delta _{p}=0$ and
\begin{equation}
\label{pd2}
\Delta _{p}=\pm\frac{1}{2} \sqrt{\Omega _{o_1}^{2}+\Omega
_{o_2}^{2}+ \Omega_{m}^{2}}\,,
\end{equation}
respectively, when $A_S=0$ at $\varphi=\pi/2$.
The widths of each spectral component can be easily
evaluated by means of the equation
\begin{equation}
\label{simx}
\chi_{\Im}/2=-\frac{N|\wp_{ad}|^{2}}{\epsilon_{0}\hbar B_S}\,,
\end{equation}
 when  $\Delta_p$ in the expression for $B_S$ (see (\ref{scabr}))
 is replaced either by zero or by the corresponding root
 of (\ref{pd2}).

\begin{figure*}[tbp]
\centering
\includegraphics[width=5.5 in]{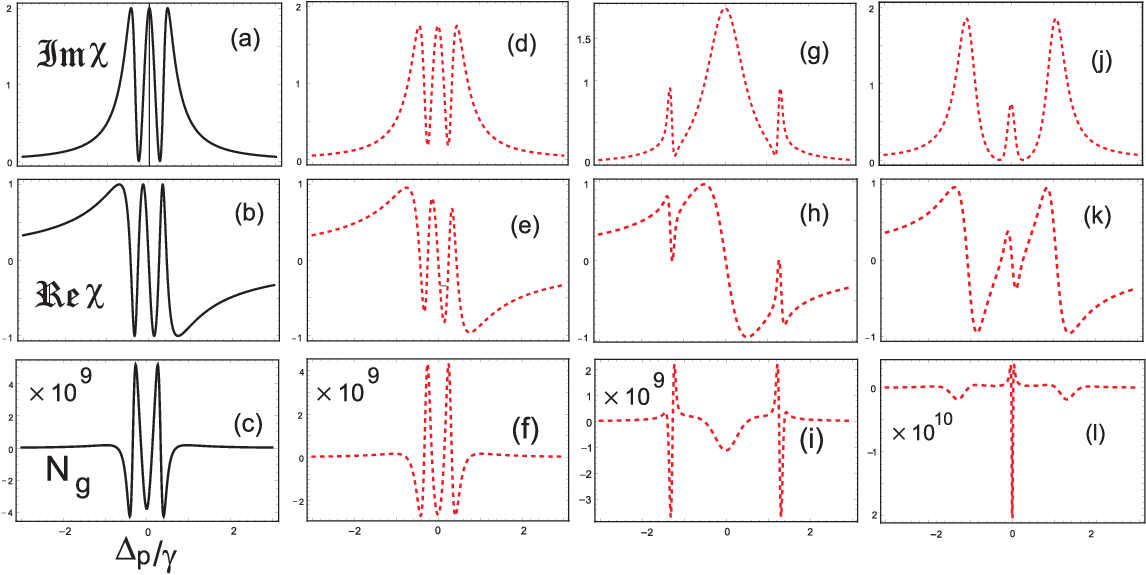}
\caption{(Color online) Absorption (a,d,g,j), dispersion
(b,e,h,k), and group index (c,f,i,l), vs the field detuning,
$\Delta_p/\protect\gamma$.
The following parameters are used: $\protect\omega=10^8%
\protect\gamma$, $V_D=0.2\protect\gamma$,
$\protect\varphi=\protect\pi/2$.
The results for equal intensities
$\Omega_{o_1}=\Omega_{o_2}=\Omega_m=0.5\protect\gamma$
are connected by solid (black) line for $\beta=0$ (a-c),
and dashed (red) line for $\beta=0.5$ (d-f).
The  results for $\Omega_m=2.5\protect\gamma$,
$\Omega_{o_1}=\Omega_{o_2}=0.5\protect \gamma$, and $\beta=0.5$
are connected by dashed (red) line (g-i).
The results for $\Omega_{o_2}=\Omega_{m}=0.5\protect\gamma$,
$ \Omega_{o_1}=2.5\protect\gamma$
are connected by dashed (red) line for $\beta=0.5$ (j-l). }
\label{f4}
\end{figure*}
Thus, we have derived the system of equations that define the real and imaginary
terms of the susceptibility in an ideal case of the one decay channel.
As before, a thorough analysis is provided for a resonant
interaction as an example, i.e., $\Delta _{i}=0,i=o_{1},o_{2},m$.

\subsection{Doppler broadening and vector mismatch}
To trace the influence of the Doppler broadening, we consider the
collinear propagation of the microwave field, two optical
and the probe fields. This consideration yields the
replacements of
$\Delta_{m}\rightarrow (\Delta _{m}+\beta kv)$,
$\Delta_{o_{1}}\rightarrow (\Delta _{o_{1}}+kv)$,
$\Delta_{o_{2}}\rightarrow (\Delta _{o_{2}}+kv)$ and
$\Delta_{p}\rightarrow (\Delta _{p}+kv)$ in (\ref{srs}),
(\ref{imsc}). The synchronization of
the microwave field with the optical driving
fields requires the condition
$\Delta_{o_{2}}=(\Delta _{o_{1}}-\Delta_{m})$ to be fulfilled
(the two photon Raman resonance condition).

Taking into account the vector mismatsh $\Delta _{m}\rightarrow
(\Delta_{m}+\beta kv)$, we use the substitution $\Delta
_{o_{2}}\rightarrow \lbrack (\Delta _{o_{1}}-\Delta
_{m})+(1-\beta)kv] $ in (\ref{srs})-(\ref{scabr}), in
addition to the substitution of the other velocity dependent
terms. The average susceptibility is defined as before
\begin{equation}
\chi (kv)=\frac{N|\wp _{ad}|^{2}}{\epsilon _{0}\hbar \Omega _{p}}[\frac{
R_{S_D}(kv)A_{S_D}(kv)+iQ_{S_D}(kv)}{A_{S_D}^{2}(kv)+B_{S_D}^{2}(kv)}\Omega
_{p}],
\end{equation}
where
\begin{eqnarray}
\label{asdd}
A_{S_{D}}(kv) &  =-2(\Delta_{o_1}+kv)\Omega_{o_{1}}^{2}+2(\Delta_{p}%
+kv)(\Omega_{o_{1}}^{2}+\Omega_{m}^{2})\nonumber\\
&  +2[[(\Delta_{o_{1}}-\Delta_{m})+(1-\beta)kv]-(\Delta_{p}+kv)]\nonumber\\
&  \times\lbrack4(\Delta_{p}+kv)(\Delta_{p}-\Delta_{o_{1}})-\Omega_{o_{2}}%
^{2}]\nonumber\\
&  +2\Omega_{o_{1}}\Omega_{o_{2}}\Omega_{m}\cos\varphi,\,\\
\label{bsdd}
B_{S_{D}}(kv) &  =\gamma\Omega_{m}^{2}+4\gamma(\Delta_{o_{1}}-\Delta_{p})\\
&
\times\lbrack(\Delta_{p}+kv)-[(\Delta_{o_{1}}-\Delta_{m})+(1-\beta
)kv]],\nonumber\\
\label{rsdd}
R_{S_{D}}(kv) &  =\Omega_{m}^{2}+4(\Delta_{o_{1}}-\Delta_{p})\\
&
\times\lbrack(\Delta_{p}+kv)-[(\Delta_{o_{1}}-\Delta_{m})+(1-\beta
)kv]]\,,\nonumber\\
\label{qsdd}
Q_{S_{D}}(kv) &  =-2A_{S_{D}}(kv)B_{S_{D}}(kv)R_{S_{D}}(kv)\,.
\end{eqnarray}

In this system the quantum interference is always dominant for
both the Doppler-free and Doppler broadened cases due to simple
losses. In fact, this a nice manifestation of the Fano-like
interference mechanism in the triplet spectroscopy. In the case of
the doublet spectroscopy, at the resonance condition, the position
of the dark line is always fixed (see e.g. Refs.\cite{zub,eit}).
In our model there are two dark lines in the absorption
spectra, independently on the strength of the driving fields at
the relative phase $\varphi=\pi/2$ (see top row of Fig.\ref{f4}).
Their positions can be changed by the intensity of
the microwave field (see (\ref{pdl})). In virtue of this fact, one
can tune up the whole system by a proper choice of the optical
fields.

\begin{figure*}[tbp]
\centering
\includegraphics[width=6in]{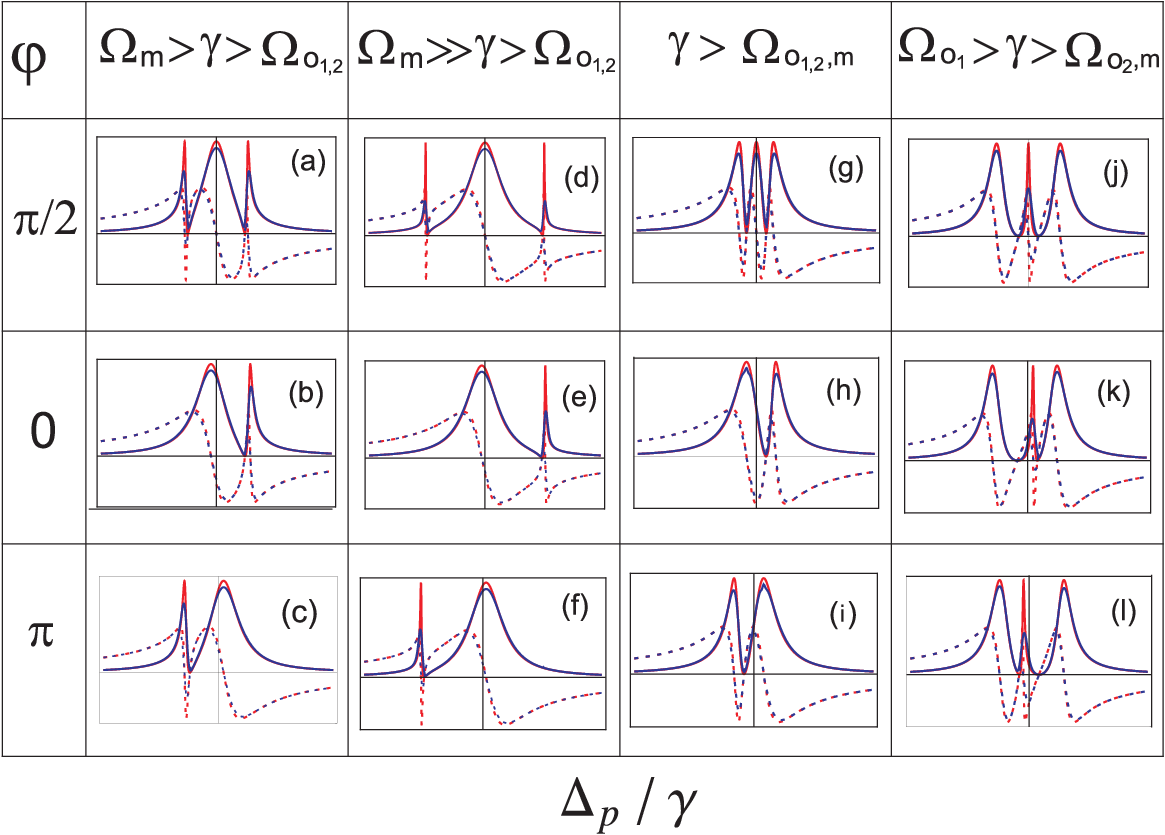}
\caption{(Color online) Absorption (solid line), dispersion
(dashed line) vs dimensionless probe field detuning
$\Delta_p/\protect\gamma$. The red lines connect the results without
Doppler broadening and vector mismatch; the blue lines
connect the results with the vector mismatch $\beta k=0.5
\protect\gamma$ and the Doppler width
$V_D=0.2 \protect\gamma$.
Panels (a,b,c): $ \Omega_{o_1}= \Omega_{o_2}=0.5\protect\gamma,
\Omega_m=1.5\protect\gamma$. Panels (d,e,f):
$\Omega_{o_1}=\Omega_{o_2=}0.5\protect\gamma, \Omega_m=3
\protect\gamma$. Panels (g,h,i):
$\Omega_{o_1}=\Omega_m=\Omega_{o_2}=0.5 \protect\gamma$. Panels
(j,k,l): $\Omega_m=\Omega_{o_2}=0.5\protect\gamma, \Omega_{o_1}=
1.5 \protect\gamma$. The other parameters are kept the same as in
Fig.6.}
\label{f7}
\end{figure*}

Evidently, the Doppler broadening degrades the system coherence.
In addition, the vector mismatch relaxes the sub- and
super-liminality (compare Figs.\ref{f4}(a-c) and
Figs.\ref{f4}(d-f)) at the equal intensity of the driving fields.
However, as it was shown in Sec.IIIB, one can control
the properties of the central absorption peak by choosing the
appropriate optical field to be dominant over other parameters of
the system. In fact, many features are simply the consequence
of the Kramers-Kronig relations. Two side and a small central peaks in
the absorption spectrum correspond to anomalous dispersions for the probe field.
The dominance of the intensity of the optical field $\Omega_{o_1}$
yields the drastic increase of the super-luminality (compare Figs.\ref{f4}f,l) and
to the attenuation of the sub-luminality of the side peaks.

At the dominance of the microwave field $\Omega_m$, for the
central broadened peak (see Fig.\ref{f4}g) the dispersion is mild
(see Fig.\ref{f4}h). The substantial enhancement in the group
index for the steep anomalous regions at the side positions of the
absorption spectra is accompanied by the attenuation of the
super-luminality at the central position due to
the flattened dispersion (compare Figs.\ref{f4}f,i).

Thus, in the proposed scheme one is able to control the
transparency windows by means of the external fields in the
triplet absorption spectroscopy. The EIT aspect related to the
absorption/dispersion properties of the proposed scheme is studied
in detail below.

\subsection{Control of the EIT windows}

To understand better the quantum interference phenomenon we
consider different relative phases $\varphi=0,\pi/2,\pi$ of the driving fields.
Note that the results for the phase
values $ \varphi=0, \pi/2$ represent a mirror reflection of those
obtained for the values $\varphi=\pi,3\pi/2$. We summarize our
findings in Fig.\ref{f7}, that are also consistent with the
results of Sec.III.

Obviously, the spectrum splits into a triplet by two dark lines
developed in the Lorentzian type absorption spectrum at $
\Delta_p=\pm\Omega_m/2$ (see Fig.\ref{f7}(g) for $\varphi=\pi/2$).
The width of each spectral component appears as $\gamma/3$ when
the intensities of the driving fields are kept equal. The quantum
interference is always dominant for the Doppler-free and Doppler
broadened cases, due to the simple probability loss. The absence of
additional radiative decay rates in the system relaxes the
influence of the vector mismatch of the microwave field.
However, if the microwave field is switched
off, the control of the EIT is reduced, since one is faced with
the traditional tripod systems (see, e.g., Ref.\cite{barry-sender}).

The double transparency windows could exist in the system as well
if one of the optical fields is switched off. In this case, the
absorption (\ref {srs}) and dispersion (\ref{imsc}) spectra become
phase insensitive, when either $\Omega_{o_1}$ or $\Omega_{o_1}$ is
set to zero in (\ref{asdd}). Therefore, to maintain phase effects
in the system, it is important to preserve  loop structure
couplings of the driving fields.

To illuminate the crucial role of the microwave
field and the relative phase of the driving fields, we neglect the
effect of the Doppler broadening and focus on the absorption
(\ref{imsc}) at the resonance condition
$\Delta_{o_{2}}=\Delta_{o_{1}}=\Delta_{m}=0$.
In this case we have
\begin{eqnarray}
A_{S_{R}} &  = &-8\Delta_{p}^{3}+2\Omega^{2}\Delta_{p}+2\Omega_{o_{1}}%
\Omega_{o_{2}}\Omega_{m}\cos\varphi,\\
R_{S_{R}}& = &\Omega_{m}^{2}-4\Delta_{p}^{2}\,,\quad
B_{S_{R}}  = \gamma R_{S_{R}}\,,\\
\Omega^{2}&=&\Omega_{o_{1}}^{2}+\Omega_{o_{2}}^{2}+\Omega_{m}^{2}.
\end{eqnarray}
We recall that the prefactor
${N|\wp_{ad}|^{2}}/({\epsilon_{0}\hbar})\equiv \gamma$ is our
natural unit. Further, for the sake of convenience we introduce
the following notations
\begin{equation}
\label{dy}
y_i=\Omega_i/2\gamma\,,\quad i=o_1,o_2, m\,, \quad x=\Delta_p/\gamma\,.
\end{equation}
Taking into account these definitions, it is straightforward to
present (\ref{imsc}) in the Fano-type form
\begin{eqnarray}
\label{sc}
&&\chi_{\Im} = \frac{(x^2-y_m^2)^2}{4F^2+(x^2-y_m^2)^2}\,,\\
\label{toe}
&&F=x^3-K^2x-M\cos\varphi\,,\\
\label{sc2}
&&K^2=(y_1^2+y_2^2+y_m^2)\,,\quad M=2y_1y_2y_m\,.
\end{eqnarray}
These equations demonstrate clearly the dependence of the
absorption on the strength of the microwave field and the
relative phase $\varphi$ of the driving fields. It appears that the
positions of the dark lines are always determined as $x=\pm
y_m\Rightarrow \Delta_p=\pm\Omega_p/2$. However, this result should
be taken cautiously, since the phase factor play important role as
well.

Evidently, the roots of  (\ref{toe}) determine the
position of three absorption maxima (see Figs.\ref{f7}a,d,g,j) in general.
In particular, at $\varphi=\pi/2$ they are defined by
(\ref{pdl},\ref{pd2}). There is, however, a significant
influence of the relative phase of the driving fields on the
tunability of the Fano-like resonances in
the triplet absorption spectrum. Two Fano-like resonances can be
tuned into a single Fano-like resonance and \textit{vise versa} in
the absorption spectrum if the relative phase of the
driving fields is varied.

Indeed, the central absorption peak (see Figs.\ref{f7}a,d,g)
merges: either with its left (right) peak at the phase $\varphi=0$
($\varphi=\pi$) at Figs. \ref{f7}b,e, (Figs.\ref{f7}c,f), or with
its right (left) peak at Fig.\ref{f7} h (Fig.\ref{f7}i).
As an example, we consider the case  Fig.\ref{f7}b:
$\Omega_{o_1}= \Omega_{o_2}=0.5\gamma, \Omega_m=1.5\gamma,
\varphi=0$. By means of (\ref{dy},\ref{sc},\ref{toe},\ref{sc2}), one readily
obtains
\begin{eqnarray}
\chi_{\Im}&=&\frac{(x^2-9/16)^2}{4(x+3/4)^2[(x-3/8)^2-17/64]^2+(x^2-9/16)^2}\nonumber\\
&=&\frac{(x-3/4)^2}{4[(x-3/8)^2-17/64]^2+(x-3/4)^2}\,.
\end{eqnarray}
The position of only one dark line is located at $x=3/4$, while
two absorption maxima are determined by $x_{1,2}=3/8\pm
\sqrt{17}/8$. In this process, one of the Fano-like resonances
(and dark lines) disappears from the spectrum. The width of the
merged peak is compensated by the increase of the width of the
corresponding side peak. In addition, for a large relative
strength, the central broadened peak is merged with the side one
(see Figs.\ref{f7}e,f).

Unlike the study in Ref. \cite{optomechanics}, where the Fano-like
resonance was tuned by using some approximation in the
opto-mechanics set-up, here the control of the single dynamical
variable $\varphi$ is enough to control the resonance profile.

Depending on the
phase and relative intensities of the driving fields, the
dispersion associated with the central and side components
of the absorption spectrum superimposes at the cost of large
absorption developed by the merging effect. Subsequently, we are
left with  one EIT window on either sides of the central line
(see Figs.\ref{f7}b,e,h and Figs.\ref{f7}c,f,i).  As a result,
the height and the width
of the absorption peak will increase, and the sub-luminal light
is converted to the super-luminal one. We expect that the probe pulse
would be greatly
attenuated during the propagation through the transparency window with
the anomalous dispersion.

Thus, the phase control can be used to
convert the double transparency behaviour of the medium to a
single transparency one and \textit{vise versa}.
In addition, at dominance one of the optical fields,
we obtain a shift of the central peak either to the left ($\phi=0$)
or to the right ($\phi=\pi$) side with the respect of its position
at ($\phi=\pi/2$) (see Figs.\ref{f7}j,k,l).

Note that the dark line positions with and without
the vector mismatch are controlled
by the intensity of the microwave field
$\Omega_m$ (see Figs.\ref{f7}a,d) as well. The larger is the intensity of
the microwave field, the widened are the transparency windows,
with a similar mirror inversion for the
dispersions. The sharper and shorter anomalous dispersions
associated with two sides sharp absorption lines are now away
from the central location of the spectrum. Therefore, the
sub-luminal probe pulse propagation through  two windows is
degraded. However, once the intensity of the optical field
$\Omega_{o_1}$, or $\Omega_{o_2}$, or both are increased, the
normal dispersions  get worsen at two windows. Subsequently,
the central absorption line becomes extremely narrowed with a giant
anomalous dispersion. In response, the slow light through the
transparency window becomes  degraded extremely (see Figs.\ref{f7}j,k,l).

We conclude that the widths of the three components of the
triplet spectrum are not dependent on the frequencies of two
optical and microwave fields.
We recall that in
the traditional one-window based EIT systems the widths of
absorption components of the corresponding spectrum are independent
on the intensities of control fields.
However, in the considered system the widths depend on their intensities
and  their coupling positions in the interaction loop relative
to the decaying energy level. For
example, the width of the central component in the considered
simplistic scheme  is narrowed
with the increase of the intensity of the optical field
$\omega_{o_1}$ or the optical field $\omega_{o_2}$, or both
of them. On the other hand, the microwave field, coupled indirectly to
the decaying energy level, produces the opposite effect.
Namely, the width of the central peak is broadened with the increase
of the microwave field intensity.
In addition, the manipulation of the EIT windows is achieved
by means of an effective tuning of
the microwave field intensity.

\section{Summary}
We have analysed thoroughly the coherence phenomena produced by three
cyclicly driven optical fields in four-level systems at zero and
weak magnetic fields. To this aim we have developed the
model in order to trace the optical response of the atomic Sodium
system, as an example. It was found that  in this system the coherence effects
yield two dark lines, in general.
This is due to the EIT phenomenon
which manifests  the interplay between the Autler-Townes and the
Fano-like mechanisms in a triplet absorption spectroscopy.

To gain a better insight into coherence phenomena we have proposed
the ideal but experimentally viable behaviour of our system at the
weak magnetic field. We have provided various sets of parameters to
explore the transition from a multiple broad-band to extremely
narrow-band windows for transparency of the probe pulse.
Such effect could exist only when two optical fields are
coupled in a combination with a microwave field to form a loop structured
interaction. In fact, we found a significant influence of
the relative phase of the driving fields on the tunability of the Fano-like
resonances in the triplet absorption spectrum.
Two Fano-like resonances in the absorption spectrum can
be tuned into a single Fano-like resonance and \textit{vise versa}.
 We have demonstrated
that the degree of the EIT (with and without the Doppler
broadening) can be controlled by altering the intensities of the external
optical fields as well.

We speculate that the controllability of the EIT windows by means
of a manipulation of the intensity of the microwave field and relative
phase of the driving fields in the considered system could help to
improve some technical aspects of devices already in use or yet to
be developed \cite{eit}.  In particular, the
coherence and interference phenomena, similar to the EIT in
gaseous media and optoelectronic materials \cite{eit,atom-vapor},
might promise some logic and functional operations such as NOT and
AND gates when a system enables to produce an enhanced double EIT
\cite{express}.

From our calculations it follows that the normal dispersions
around the two dark lines are extremely flattened. As a result,
the sub-luminality is attenuated (see Fig.\ref{f4}j,k,l). As a
matter of fact, it is the microwave field enables to one to
control the positions of the dark lines in the triplet
spectroscopy, and, subsequently, to change the sub- and
super-luminality of the probe pulse.

Steep anomalous dispersions at the corresponding windows are
accompanied with a relatively small but sharp absorption. It
results in a drastic enhancement of the super-luminal group
velocity of the probe pulse. The enhanced group velocity of light
promises various experimental applications. For example, the
quality of an image measured by a light pulse with a relatively
low group velocity could be improved if measured by a
super-luminal light pulse with a significantly high group velocity
\cite{image}.

\section*{Acknowledgement}

F.G. is grateful for the warm hospitality at JINR.
This work was supported in part by Russian Foundation for Basic Research, Grant 14-02-00723,
 and by COMSATS
Institute of Information Technology.

\end{document}